%% file: NewPaper_IV.tex
\documentclass[12pt]{article}
\usepackage{amsmath,amssymb,yfonts,esint}

\usepackage{wrapfig}
\usepackage{floatflt}
\usepackage{mathrsfs}

\usepackage[pdftex]{color,graphicx}

\usepackage{dsfont}

\newcommand{\be}{\begin{equation}}
\newcommand{\ba}{\begin{eqnarray}}
\newcommand{\ea}{\end{eqnarray}}
\newcommand{\nn}{\nonumber}

\def\d{\delta}
\def\e{\epsilon}

\def\m{\mu}
\def\n{\nu}

\def\s{\sigma}
\def\t{\tau}

\def\G{\Gamma}

\def\OO{\Omega}

\def\ca{{\cal A}}
\def\cb{{\cal B}}

\def\ch{{\cal H}}

\def\co{{\cal O}}
\def\cp{{\cal P}}

\newcommand{\pa}{\partial}

\newcommand{\bbC}{{\Bbb C}}

\fontfamily{yfrak}

\begin{document}

\vskip 15mm

\begin{center}

{\Large\bfseries On Semi-Classical States of Quantum Gravity\\[1ex] and Noncommutative Geometry 
}

\vskip 4ex

Johannes \textsc{Aastrup}$\,^{a}$\footnote{email: \texttt{johannes.aastrup@uni-muenster.de}},
Jesper M\o ller \textsc{Grimstrup}\,$^{b}$\footnote{email: \texttt{grimstrup@nbi.dk}},\\ Mario \textsc{Paschke}$^{a}$\footnote{email: \texttt{mario.paschke@uni-muenster.de}} \& Ryszard \textsc{Nest}\,$^{c}$\footnote{email: \texttt{rnest@math.ku.dk}}

\vskip 3ex  

$^{a}\,$\textit{Mathematical Institute, University of M\"unster,\\ Einsteinstrasse 62, D-48149 M\"unster, Germany}
\\[3ex]
$^{b}\,$\textit{The Niels Bohr Institute, University of Copenhagen, \\Blegdamsvej 17, DK-2100 Copenhagen, Denmark}
\\[3ex]
$^{c}$ \textit{Mathematical Institute, University of Copenhagen,\\ Universitetsparken 5, DK-2100 Copenhagen, Denmark}
\end{center}

\vskip 3ex

\begin{abstract}

We construct normalizable, semi-classical states for the previously proposed model of quantum gravity which is formulated as a spectral triple over holonomy loops. 
The semi-classical limit of the spectral triple gives the Dirac Hamiltonian in 3+1 dimensions. 
Also, time-independent lapse and shift fields emerge from the semi-classical states. Our analysis shows that the model might contain fermionic matter degrees of freedom. 

The semi-classical analysis presented in this paper does away with most of the ambiguities found in the initial semi-finite spectral triple construction. The cubic lattices play the role of a coordinate system and a divergent sequence of free parameters found in the Dirac type operator is identified as a certain inverse infinitesimal volume element.

\end{abstract}

\newpage
\tableofcontents

\section{Introduction }

A critical test of any quantum model is the existence of a semi-classical limit. This limit - its existence once established - should make contact to known physics, explain qualitative and quantitative results, and thereby render credibility to the model. Most importantly, the semi-classical limit serves to confirm the operational interpretation of the observables of the model. Furthermore, as there exist infinitely many inequivalent quantizations of classical field theories, the semi-classical limit often provides an important tool to distinguish physical relevant models.

The semi-finite spectral triple over a configuration space of connections constructed in  \cite{Aastrup:2005yk} -
 \cite{Aastrup:2009ux} constitute a non-perturbative quantum model. The spectral triple emerges from a fusion between noncommutative geometry \cite{ConnesBook,Connes:1996gi} and canonical quantum gravity \cite{Thiemann:2001yy}-\cite{Ashtekar:2004eh}. It involves an algebra of holonomy loops and a Dirac type operator that resembles a global functional derivation operator. Its existence - as a mathematical entity - was established in  \cite{Aastrup:2008wb,Aastrup:2008zk}. Its interpretation in terms of a non-perturbative quantum field theory is immediate since the interaction between the algebra and the Dirac type operator reproduces the Poisson bracket of general relativity, formulated in terms of Ashtekar variables, and of Yang-Mills theory.

What remained unresolved, in the papers  \cite{Aastrup:2005yk} -
 \cite{Aastrup:2009ux}, was the exact physical interpretation of the spectral triple construction. It was not clear whether the model should be understood in terms of gravity or Yang-Mills theory, or something else. In particular, no substantial results concerning a semi-classical limit were obtained.

In this paper we make the first steps towards a semi-classical analysis. Drawing on results by Hall \cite{Hall1,Hall2} concerning coherent states on compact Lie-groups, we construct semi-classical states over the configuration space of connections. This analysis enlightens us on two fronts:

First, at a conceptual level, the semi-classical analysis entails a clearer physical interpretation of the semi-finite spectral triple. In particular, we find that the Dirac type operator descents,  in this limit, to a Dirac Hamiltonian on a 3+1 dimensional ultra-static space-time. Through a careful analysis of the Poisson structure of general relativity we first obtain an interpretation of the constituents of the Dirac type operator as quantized triad field operators. In short, the Dirac type operator appears as an infinite sum of quantized triad field operators. In the semi-classical limit, these triad operators entail classical triad fields which appear in the classical Dirac operator. The special class of semi-classical states constructed in this paper suggest an interpretation as one-fermion states for a spinor field on the ultra-static space-time. This interpretation has, however, a problem since the scalar product induced on this space depends on the choosen coordinates. 
Nevertheless, we believe that our analysis indicates that the semi-finite spectral triple should be understood in terms of quantum gravity coupled to quantized matter fields. Indeed, if the time-scale is chosen appropriately, then the scalar product becomes coordinate independent.

Second, at a more technical level, the semi-classical analysis resolves several questions and ambiguities concerning the construction of the semi-finite spectral triple. For instance, the triple is build over a countable system of nested graphs. In \cite{Aastrup:2008zk}  it was clear that the construction would work for a large class of such systems of  graphs and no mechanism was found to single out one system of graphs from another. Furthermore, it was also clear that two spectral triples, based on different systems of graphs, would constitute entirely different models. This ambiguity is resolved through the semi-classical analysis: we find that a system of {\it cubic lattices} is singled out as "natural" with an interpretation as a choice of a coordinate system. This coordinate system is made to coincide with the coordinate system used to write down the Ashtekar variables and their Poisson bracket.

Moreover, the construction of the Dirac type operator involves an infinite series of free parameters which is required to diverge in order for the operator to have a compact resolvent. In the papers \cite{Aastrup:2005yk} -
 \cite{Aastrup:2009ux} no clear physical interpretation of these parameters were found. Again, the semi-classical analysis resolves this ambiguity: it identifies the series of free parameters as the inverse infinitesimal, Euclidean volume element, the divergence arising through a continuum limit where the volume elements approach zero.

Clearly, the introduction of finite graphs breaks diffeomorphism invariance. In loop quantum gravity \cite{Thiemann:2001yy}-\cite{Ashtekar:2004eh}, which is also based on an inductive system of graphs \cite{Ashtekar:1993wf}-\cite{Ashtekar:1996eg}, the philosophy is to include all\footnote{to be precise, all piece-wise analytic graphs.} possible graphs and thereby restore the symmetries in the inductive limit of graphs and Hilbert spaces. This renders the limiting Hilbert space non-separable, something which probably obstructs the construction of a spectral triple \cite{Aastrup:2005yk}. In this paper we find that the constructed semi-classical limit does not depend on finite parts of the inductive system of lattices. Thus, in this limit the lattices seemingly dissapear and the symmetries, broken by the initial choice of graphs, are restored. This means that the expressions for the classical Dirac operator and the Dirac Hamiltonian, found in the semi-classical limit, are coordinate covariant. 

The finding that cubic lattices are singled out by the semi-classical analysis plays well with recent results by Flori and Thiemann which state that, in loop quantum gravity, {\it only} lattices with cubic topology give the right semi-classical limit  \cite{Flori:2008nw}.\\

This paper is organized as follows: In section 2 we briefly review noncommutative geometry and Connes work on the standard model. In section 3 we introduce Ashtekar variables together with their dual variables, the loop and flux variables. In section 4 we then review the construction of the semi-finite spectral triple. First, a spectral triple is constructed on a fixed graph, and subsequently a continuum limit of spectral triples is taken over an infinite system of ordered graphs. In section 5 we comment on the underlying space of generalized connections and section 6 is concerned with a careful analysis of the relationship between the spectral triple construction and the Poisson bracket between flux and loop variables. Finally, section 7 is concerned with the semi-classical states. In section 8 we give a conclusion.

\section{Noncommutative geometry }

It is a central observation in noncommutative geometry, due to Connes, that the metric of a compact manifold can be recovered from the Dirac operator together with its interaction with the smooth functions on the manifold \cite{ConnesBook}. In other words the metric is completely determined by the triple
$$(C^\infty (M),L^2(M,S), D)\;.$$
This observation leads to a noncommutative generalization of Riemmanian geometries. Here the central objects are {\it spectral triples} $(A,H,D)$, where $A$ is a not necessarily commutative algebra; $H$ a Hilbert space and $D$ an unbounded selfadjoint operator called the Dirac operator. The triple is required to satisfy some interplay relations between $A,H,D$ mimicking those of $(C^\infty (M),L^2(M,S), D)$.  The choice of the Dirac operator $D$ is strongly restricted by these
requirements. 

In physics, a key example of a noncommutative geometry comes from particle physics. Again, it was Connes who realized that the entire data of the standard model coupled to general relativity can be understood as a single, gravitational model formulated in terms of a spectral triple \cite{Connes:1996gi}\cite{Chamseddine:1991qh}-
\cite{Chamseddine:2007ia}. Here, the algebra is an {\it almost commutative algebra}
$$
A=C^\infty(M)\otimes A_F\;,
$$
where $A_F$ is the algebra $\mathbb{C}\oplus\mathbb{H}\oplus M_3(\mathbb{C})$. The corresponding Dirac operator then consists of two parts,
$$ D= D_M + D_F \;, $$
one of which is the standard Dirac operator $D_M$ on $M$. 
The other part, $D_F$, is given by a matrix-valued function on the manifold $M$, that encodes the metrical aspects of the states over the algebra $A_F$. 
It is a highly nontrivial and very remarkable fact that the above mentioned requirements  for  Dirac operators 
force $D_F$ to contain the non-abelian gauge fields of the standard model and the Higgs-field together with their couplings to the elementary 
fermion fields. In particular the Higgs-field thus obtains a geometrical interpretation as being a part of the gravitational field on a noncommutative space.
Even more so, the classical action of the standard model coupled to the Einstein-Hilbert action, in the Euclidean signature, emerges from the spectral triple through the so-called spectral action principle \cite{Chamseddine:1991qh}, which states that physics only depends on the spectrum of the Dirac operator. 

In view of the widely held opinion that quantum effects of the gravitational field will necessarily lead to a noncommutativity of spacetime this observation
indicates that the gauge interactions and the appearance of the  Higgs field may be interpreted as quantum effects of the gravitational interactions.
In other words they are the first shadows of the noncommutativity of spacetime, visible at the length  scale corresponding to the $Z$-mass. 

It should be mentioned in this respect that the spectral action does not directly reproduce the correct coupling constants of the standard model.
In fact it only allows for lesser free parameters than the standard model. 
In order to obtain the measured coupling constants for the electromagnetic and strong interactions to a fairly good approximation,
Connes and Chamseddine applied renormalization group methods in 
\cite{Chamseddine:1996rw} and subsequent publications.  This  analysis ultimately leads to a prediction of the Higgs mass \cite{Chamseddine:2006ep}. The predicted value, which was based on the assumption of "the big dessert", was recently excluded by Tevatron data.  Nevertheless it is very remarkable that
the use of quantum field theoretical concepts is absolutely essential  here to obtain a physically reasonable classical action.  

To our point of view this strongly indicates that the spectral triple used by Connes ad Chamseddine should be viewed as the semi classical low energy limit of some genuine quantum theory. One may then also hope that other quantum corrections present in the full theory provide a more realistic value for the  
Higgs mass.
Since the noncommutative describtion of the standard  model is entirely gravitational this full theory should, presumably, be a theory of the quantized gravitational field.
Thus, if there were already a theory of quantum gravity one should certainly investigate whether it admits some semi-classical states that resemble this almost commutative spectral triple.
 
It was these considerations which motivated the construction of the semi-finite spectral triple over a configuration space of connections \cite{Aastrup:2005yk} -
 \cite{Aastrup:2009ux} . The idea is to seek a general framework which combines the machinery and ideas of noncommutative geometry with elements of quantum gravity. The final goal, then, is to make contact to Connes work on the standard model through the formulation of a semi-classical analysis.

\section{Ashtekar variables and holonomy loops}
\label{canonicalgravity}


We start with some notation. Let $M$ be a 4-dimensional globally hyperbolic manifold with a vierbein $E_\m^A$ and a space-time metric $G_{\m\n}= E_\m^A E_\n^B\eta_{AB}$ where $\eta_{AB}=\mbox{diag}(-1,1,1,1)$ is the corresponding tangent space metric. Here the letters $\m,\n,...$ and $A,B,...$ denote curved and flat space-time indices respectively. Next, take a foliation of $M$ according to $M = \mathbb{R}\times \Sigma$ where $\Sigma$ is a spatial manifold. Let $g_{mn}=e_m^a e_{na}$ be the corresponding spatial metric and $e_m^a$ the spatial dreibein. Here the letters $m,n, ...$ and $a,b,...$ denote curved and flat spatial indices.

The Ashtekar variables \cite{Ashtekar:1986yd,Ashtekar:1987gu} consist first of a complex $SU(2)$ connection $A_{m}^a(x)$ on $\Sigma$. The Ashtekar connection is a certain complex linear combination of the spatial spin connection and the extrinsic curvature of $\Sigma$ in $M$. The canonically conjugate variable to $A_{m}^a(x)$ is the inverse densitised dreibein
\[
\bar{E}^m_a = e e_a^m\;,
\]
where $e=\mbox{det}( e_m^a)$. This set of variables satisfy the Poisson bracket
\[
\{ A_{m}^a(x),\bar{E}_b^n(y)\} =  \kappa\d^a_b\d_m^n \d^{(3)}(x,y)\;,
\]
where $\kappa$ is the gravitational constant. The formulation of canonical gravity in terms of connection variables permits a shift to loop variables which are taken as the holonomy transform  
\[
h_l(A)  = \cp \mbox{exp}\int_l A_m dx^m\;,
\]
along a loop $l$ in $\Sigma$. To define a conjugate variable to $h_l(A)$ let $dF_a$ be the flux of the triad field $\bar{E}_a^m$ corresponding to an infinitesimal area element of the spatial manifold $\Sigma$, which can be written
\begin{equation}
dF_a = \e_{mnp} \bar{E}^m_a dx^n\wedge dx^p \;.
\nonumber
\end{equation}
Given a 2 dimensional surface $S$ in $\Sigma$ we write the total flux of $\bar{E}_a^m$ through $S$ 
\[
F_{S}^a= \int_S dF^a\;.
\]
Next, consider a surface $S$ and let $l=l_1\cdot l_2$ be a line segment in $\Sigma$ which intersect $S$ at the point $l_1\cap l_2$. The Poisson bracket between the flux and holonomy variables read \cite{Thiemann:2001yy}
\begin{equation}
\{ h_l, F_S^a \} =  \iota (S,l) \kappa h_{l_1}\t_a h_{l_2}\;.
\label{Poisson}
\end{equation}
where $\t$ denote the generators of the Lie algebra of $G$. Here, $\iota$ is given by
\[
\iota(S,l) = \pm 1, 0
\]
depending on the intersection between $S$ and $l$.\\


\section{Spectral triples of holonomy loops}

In this section we outline the construction of the semi-finite spectral triple first presented in \cite{Aastrup:2008wa,Aastrup:2008wb} and further developed in \cite{Aastrup:2008zk}.
This spectral triple combines ideas and techniques of canonical gravity and noncommutative geometry. 
We first construct a spectral triple at the level of a finite graph. Next we take the limit of such spectral triples, over an infinite system of ordered graphs, to obtain a  limiting spectral triple.

\subsection{Holonomy loops}

\begin{figure}[t]
\begin{center}
\resizebox{!}{3cm}{
 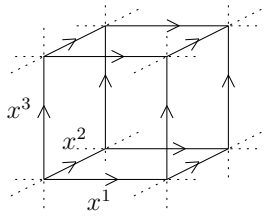}
\end{center}
\label{plaquet}
\caption{A plaquet in the lattice $\G$.}
\end{figure}
Let $\G$ be a 3-dimensional, finite, cubic lattice. Let $\{v_i\}$ and $\{l_j\}$ denote vertices and edges in $\G$, respectively. The edges in $\G$ are oriented according to the three main directions in $\G$, the $x^1$- $x^2$- and $x^3$-directions, see figure 1. Thus, an edge $l$ is a map
\[
l:\{0,1\}\rightarrow \{v_i\}\;,
\]
where $l(0)$ and $l(1)$, the start and endpoints of $l$, are adjacent vertices in $\G$. A sequences of edges $\{ l_{i_1}, l_{i_2},\ldots, l_{i_n}\}$ where $l_{i_j}(1)= l_{i_{j+1}}(0)$ is a based loop if $l_{i_1}(0)=l_{i_n}(1)=v_0$ where $v_0\in \{v_i\}$ is a preferred vertex in $\G$ called the basepoint. An edge has a natural involution given by reversing its orientation. Thus, 
\[
l_i^*(t)=l_i(1-t)\;,
\]
and the involution of a loop $L=\{ l_{i_1}, l_{i_2},\ldots, l_{i_n}\}$ is given by
\[
L^*=\{ l^*_{i_n}, \ldots,l^*_{i_2}, l^*_{i_1}\}\;.
\]
In the following we shall discard trivial backtracking which means that we introduce the equivalence relation
\[
\{\ldots,l_{i_{j-1}}, l_{i_j}, l^*_{i_j},\ldots\} \sim \{\ldots, l_{i_{j-1}},\ldots\}\;,
\]
and let a loop $L$ be an equivalence class with respect hereto. 

The product between two loops $L_1=\{l_{i_j}\}$ and $L_2=\{l_{i_l}\}$ is simply given by gluing the loops to form a new sequence of edges:
\[
L_1\cdot L_2 = \{l_{i_1},\ldots,l_{i_n},l_{k_1},\ldots,l_{k_m}\}\;.
\]
One easily checks that the involution equals an inverse which gives the set of loops in $\G$ the structure of a group.  

Finally, we consider finite series of loops
\begin{equation}
a =\sum_i a_i L_i\;,\quad a_i\in \mathbb{C}\;,
\label{form}
\end{equation}
with the involution
\[
a^* = \sum_i  \bar{a}_i L^*_i\;,
\]
and the product  between $a$ and a second element $b=\sum_j b_j L_j$ 
\[
a \cdot b = \sum_{i,j} (a_i b_j) L_i\cdot L_j \;.
\]
The set of elements of the form (\ref{form}) is a $\star$-algebra. We denote this algebra by $\cb_\G$.

\subsection{ Generalized connections}

Next, let $G$ be a compact, connected Lie-group.  For the aim of this paper it is natural to choose $G=SU(2)$. We shall, however, develop the formalism for  general groups. Let $\nabla$ be a map
\[
\nabla:\{l_i\}\rightarrow G\;,
\]
which satisfies
\[
 \nabla(l_i) = \nabla(l_i^*)^{-1}\;,
\]
and denote by $\ca_\G$ the set of all such maps. Clearly,
\[
\ca_\G \simeq G^{n_\G}\;,
\]
where the total number of vertices in $\G$ is written $n_{\G}$. 
Given a loop $L=\{ l_{i_1}, l_{i_2},\ldots, l_{i_n}\}$ let
\[
\nabla(L)= \nabla(l_{i_1})\cdot \nabla(l_{i_2})\cdot \ldots\cdot \nabla(l_{i_n})\;.
\]
This turns $\nabla$ into a homomorphism from the hoop group into $G$ and provides a norm on $\cb_\G$
\[
\parallel a\parallel = \sup_{\nabla\in\ca_\G}\parallel\sum_i a_i\nabla(L_i)\parallel_{G}\;,\quad    a\in\cb_{\G}\;,
\]
where the norm on the rhs is the matrix norm given by a choosen representation of $G$. The closure of the $\star$-algebra of loops with respect to this norm is a $C^\star$-algebra\footnote{Note that the natural map from $\cb_\G$ to $B_\G$ is not necessarily injective.}. We denote this loop algebra by $B_\G$.\\

\subsection{A spectral triple over $\ca_\G$}

First, let $\ch_\G$ be the Hilbert space
\[
L^2(G^{n_\G},Cl(T^*G^{n_\G})\otimes M_l(\mathbb{C}))\;,
\]
where $L^2$ is with respect to the Haar measure and where $l$ is the size of the matrix representation of $G$. Here, $Cl(T^*G^{n_\G})$ is the Clifford bundle of the cotangent bundle over $G^{n_\G}$ with respect to a chosen left and right invariant metric.  There is a natural representation of the loop algebra on $\ch_\G$ given by
\[
f_L \cdot \Psi(\nabla)= (1\otimes \nabla(L))\Psi(\nabla)\;,\quad \Psi\in \ch_\G\;,
\]
where the first factor acts on the Clifford bundle and the second factor acts on the matrix factor in $\ch_\G$.

Next, denote by $D_\G$ a Dirac operator on $\ca_\G$. The precise expression for $D_\G$ will be determined below through the process of taking the continuum limit of the construction. $D_\G$ acts on the factor of $\ch_\G$ which involves the Clifford bundle.
In total, the triple $(B_\G,\ch_\G,D_\G)$  is a geometrical construction over $\ca_\G$.

\subsection{The limiting spectral triple}

\begin{figure}[t]
\begin{center}
\resizebox{!}{2.5cm}{
 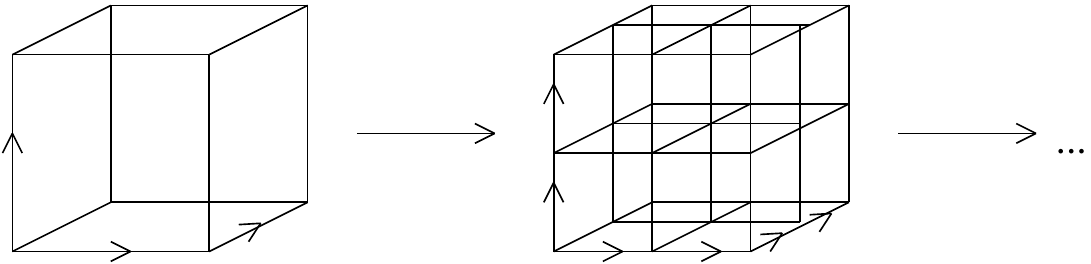}
\end{center}
\caption{Subdivision of a cubic lattice cell into 8 new cells.}
\end{figure}
The goal is to obtain a spectral triple over the space $\ca$. To do this we take the limit of spectral triples over the intermediate spaces $\ca_{\G}$.\\

Let $\{\G_i\}$, $i\in \mathbb{N}$, be an infinite sequence of 3-dimensional, finite, cubic lattices where $\G_{i+1}$ is the lattice obtained from $\G_i$ by subdividing each elementary cell in $\G_i$ into 8 new cells. This process involves the subdivision of each edge $l_j$ in $\G_i$ into two new edges in $\G_{i+1}$ together with the addition of new vertices and edges, see figure 2. We denote the initial lattice by $\G_0$.
Corresponding to this sequence of cubic lattices there is a projective system $\{\ca_{\G_i}\}$ of spaces obtained from the graphs $\{\G_i\}$, together with natural projections between these spaces
\begin{equation}
P_{i,i+1}: \ca_{\G_{i+1}}\rightarrow \ca_{\G_{i}}\;.
\label{structure}
\end{equation}
Consider now a system of triples
\[
(B_{\G_i},\ch_{\G_i},D_{\G_i})\;,
\]
with the restriction that these triples are compatible with the projections (\ref{structure}). This requirement is easily satisfied for the algebras and the Hilbert spaces, see \cite{Aastrup:2008wb}. For the Dirac type operators, however, some care must be taken. The problem reduces to the simple case where an edge in $\G_i$ subdivided into two edges in $\G_{i+1}$, see figure 3.1, which corresponds to the projection
\begin{figure}[t]
\begin{center}
\resizebox{!}{3cm}{
 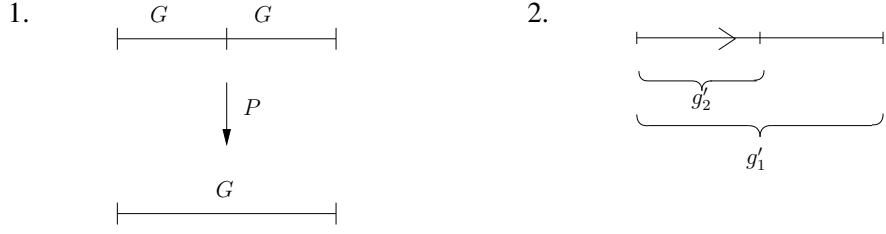}
\end{center}
\caption{A subdivision of an edge into two and the new parameterization of the edge.}
\end{figure}
\begin{equation}
P: G^2 \rightarrow G\;,\quad (g_1,g_2)\rightarrow g_1\cdot g_2\;,
\label{firstproj}
\end{equation}
and a corresponding map between Hilbert spaces
$$
P^*: L^2(G,Cl(T^*G)\otimes M_l)\rightarrow L^*(G^2,Cl(T^*G^2)\otimes M_l)\;.
$$
The compatibility condition for the Dirac type operator reads
\begin{equation}
P^* (D_{1} v)(g_1,g_2) = D_{2} (P^*v)(g_1,g_2) \;,\quad v\in L^2(G,Cl(T^*G)\otimes M_l)\;.
\nonumber
\end{equation}
Here $D_{1}$ is the Dirac operator on $G$,  and $D_{2} $ is the corresponding Dirac operator on $G^{2}$. 

Consider the following change of variables
\begin{equation}
\Theta: G^2\rightarrow G^2\; ;   (g_1,g_2)\rightarrow (g_1\cdot g_2,g_1)=: (g_1',g_2')\;,
\label{change}
\end{equation}
for which projection (\ref{firstproj}) obtains the simple form
\begin{equation}
P(g_1',g_2') = g_1'\;.
\label{proJ2}
\end{equation}
This change of variables corresponds to a new parameterization of the edge, see figure 3.2.
It is now straight forward to write down a Dirac operator on $G^2$ which is compatible with the projection (\ref{proJ2}). Basically, we can pick any Dirac operator of the form
\begin{equation}
D_2  = D_1 + a D'_2\;,\quad a\in \mathbb{R}\;,
\nonumber
\end{equation}
where $D'_2$ is a Dirac operator on the copy of $G$ in $G^2$ whose coordinates are eliminated by the projection (\ref{proJ2}). At this point the choice of the operator $D'_2$ is essentially unrestricted with $a$ being an arbitrary real parameter. However, for reasons explained in \cite{Aastrup:2008zk} it turns out that  $D_1$ and $D'_2$ should of the form
\begin{equation}
D_i = \sum_j    {\bf e}_i^j \cdot L_{{\bf e}_i^j}\;,
\label{dir}
\end{equation}
where the product is Clifford multiplication. In equation (\ref{dir}) $\{e_i^j\}$ denotes a left-translated orthonormal basis of $T^*G$ where $G$ is the $i$'th copy in $G^n$. $L_{e_i^j}$ denotes the corresponding differential. For later reference we denote by $R_{e_i^j}$ the right translated vector fields.

This line of analysis is straightforwardly generalized to repeated subdivisions. At the level of the $n$'th subdivision of the edge the change of variables which generalizes (\ref{change}) reads
\begin{eqnarray}
\Theta:G^n  \rightarrow G^n\; ; &&
\nn\\ (g_1,g_2,\ldots,g_n)&\rightarrow& (g_1\cdot g_2 \cdot \ldots\cdot g_n, g_2 \cdot \ldots\cdot g_n, \ldots, g_n)
\nn\\&:=&(g_1',g_2',\ldots,g_n')
\label{genchange}
\end{eqnarray}
which corresponds to the structure maps
\[
P_{n,n/2}: G^n \rightarrow G^{n/2}\; ; \quad (g_1',g_2',g_3',\ldots,g_n') \rightarrow (g_1',g_3',\ldots,g_{n-1}')\;.
\]
Again, it is straightforward to construct a Dirac type operator compatible with these structure maps. This construction gives rise to a series of free parameters $\{a_i\}$, one for each subdivision. 
Thus, by solving the $G^2\rightarrow G$ problem repeatedly, and by piecing together the different edges, we end up with a Dirac type operator on the level of $\Gamma_n$ of the form
\begin{equation}
D_{\G_n}=\sum_i a_i D_i\;,
\label{general}
\end{equation}
where $D_i$ is a Dirac type operator corresponding to the $i$'th level of subdivision in $\ca_{\G_n}$.

\begin{figure}[t]
\begin{center}
\resizebox{!}{3.5cm}{
 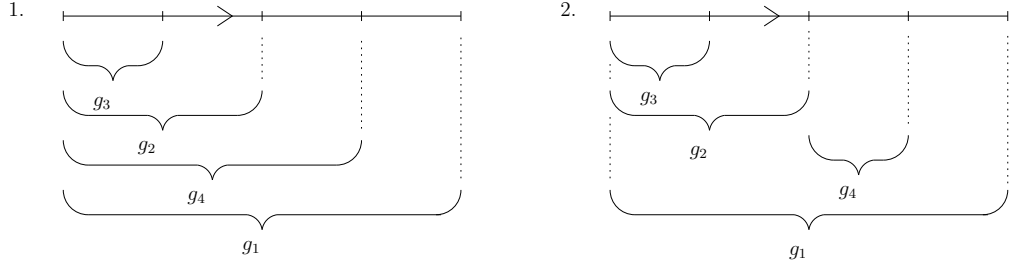}
\end{center}
\label{AATWO}
\caption{Two different types of partition which will lead to different Dirac type operators. The second partition is the one which we will later argue is "natural".}
\end{figure}
The change of variables in (\ref{genchange}) is the key step to construct $D_{\G_n}$. However, there will be many different partitions of the line segment which simplify the structure maps and lead to different Dirac type operators, see figure 4. This ambiguity was also commented on in \cite{Aastrup:2008zk}. In subsequent sections we will argue that a single type of subdivision stand out as "natural" due to the classical interpretation of the corresponding Dirac type operator.\\

We are now ready to take the limit of the triples $(B_{\G_i},\ch_{\G_i},D_{\G_i})$. First, the Hilbert space $\ch$ is the inductive limit of the intermediate Hilbert spaces $\ch_{\G_i}$. That is, it is constructed by adding all the intermediate Hilbert spaces 
$$
\ch' =\oplus_{\G\in \{\G_i\}}L^2 (G^{n(\G)},Cl(T^*G^{n(\G)})\otimes M_l({C}))/N\;, 
$$
where $N$ is the subspace generated by elements of the form 
$$
(\ldots , v, \ldots , -P^*_{ij}(v),\ldots )\;,
$$
where $P^*_{ij}$ are the induced maps between Hilbert spaces. The Hilbert space $\ch$ is then the completion of $\ch'$. The inner product on $\ch$ is the inductive limit inner product. This Hilbert space is manifestly separable. 
Next, the algebra
$$
\cb:= \lim_{\stackrel{\G}{\longrightarrow}}\cb_{\G}\;
$$
contains loops defined on a simplicial complex $\G_n$ in $\{\G_n\}$. 
Finally, the Dirac-like operator $D_{\G_n}$ descends to a densely defined operator on the limit Hilbert space $\ch$
$$
D = \lim_{\stackrel{\G}{\longrightarrow}}D_{\G_n}\;.
$$
We factorize $\ch'$ in 
$$\lim L^2(G^{n(\Gamma)}, M_l)\otimes \lim Cl(T^*_{id}(G^{n(\Gamma)})).$$ 
On $\lim Cl(T^*_{id}(G^{n(\Gamma)}))$ there is an action of the algebra $\lim Cl(T^*_{id}(G^{n(\Gamma)}))$. The completion of this algebra with respect to this action is the CAR algebra and admits a normalized trace, i.e. $tr(1)=1$. Let $Tr$ be the ordinary operator trace on the operators on $\lim L^2(G^{n(\Gamma)}, M_l)$ and define $\tau =Tr\times tr$. In \cite{Aastrup:2008wb} we prove that for a compact Lie-group $G$ the triple $(\cb,\ch,D)$ is a semi-finite spectral triple with respect to $\tau$  when the sequence $\{a_n\}$ converges to infinty. This means that:
\begin{enumerate}
\item
$
(1+D^2)^{-1}
$
is $\tau$-compact, i.e. can be approximated in norm with finite trace operators, and
\item
the commutator $[D,a]$ is bounded.
\end{enumerate}

\section{The space of connections}

Let us now turn to the spaces $\ca_{\G_i}$ and their projective limit.
Denote by
\[
\overline{\ca} := \lim_{\stackrel{\G}{\longleftarrow}}\ca_\G\;.
\]
Further, given a trivial principal $G$-bundle denote by $\ca$ the space of all smooth connections herein. In \cite{Aastrup:2008wb} we prove that $\ca$ is densely embedded in $\overline{\ca} $:
\[
\ca \hookrightarrow \overline{\ca} \;.
\]
This fact justifies the terminology {\it generalized connections} for the completion $\overline{\ca}$ and shows that the semi-finite spectral triple $(\cb,\ch,D)$ is indeed a geometrical construction over the space $\ca$ of smooth connections.

\section{The quantization of the Poisson bracket}
\label{ThePoissonStructure}

To determine the relation between the construction of the spectral triple $(\cb,\ch,D)$ and the formulation of canonical gravity in terms of loop and flux variables satisfying the Poisson bracket (\ref{Poisson}), we calculate the commutator between the Dirac type operator $D$ and an element in the loop algebra $\cb$. Consider first a single line element $l_i$ and the corresponding group element $ \nabla(l_i)\in G$. We assume that the copy of $G$ in $\ca_{\G_n}$ assigned to $l_i$ corresponds to the $m$'th subdivision of the initial cubic lattice.  We then find
\begin{equation}
[D, \nabla(l_i)]  = a_m \sum {\bf e}_i^k \cdot \nabla(l_i) \sigma^k \;,
\nonumber
\end{equation}
where $\sigma^k$ are generators in the Lie-algebra $\mathfrak{g}$.  Also, consider a loop $L=\{ l_{i_1},  l_{i_2},\ldots,  l_{i_n} \}$ and the commutator
\[
[D, f_{L}] = [D, \nabla(l_1)] \cdot \nabla(l_2)\ldots \nabla(l_n) + \nabla(l_1)[D, \nabla(l_2)] \ldots \nabla(l_n) + \ldots 
\]
These formula show that a commutator between $D$ and an element of the algebra $\cb$ inserts Lie-algebra generators at vertices in the graphs $\{\G_i\}$. This general structure is similar to the structure of the Poisson bracket (\ref{Poisson}) and suggest that the interaction between the Dirac type operator $D$ and the loop algebra $\cb$ is related to a representation of the Poisson bracket (\ref{Poisson}). 

\begin{figure}[t]
\begin{center}
\resizebox{!}{2.5cm}{
 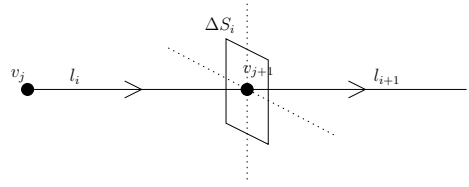}
\end{center}
\caption{The surface $\Delta S_i$.}
\end{figure}
Consider again a single edge $l_i$ which we now for simplicity assume to belong to the initial lattice. Let $l_i(0)=v_j$ and $l_i(1)=v_{j+1}$ where $v_j$ and $v_{j+1}$ are vertices in $\G_0$. Let us also assume that $l_i$ runs in the $x^1$-direction in $\G_0$. Also, let $\nabla(l_i)$ belong to the $i$'th copy of $G$ in $\ca_{\G_0}$. The commutator between the left-invariant vector field ${\bf e}^a_i$ and the group element $\nabla(l_i)$ gives
\[
[L_{{\bf e}^a_i},\nabla(l_i)] =   \nabla(l_i) \sigma^a \;.
\]
This shows that $L_{{\bf e}^a_i}$ corresponds to a quantization of a flux variable $F^a_S$ where the surface $S$ intersects $l_i$ at $v_{j+1}$. Actually, the surface $S$ is of no significance here except for its intersection point with the vertex $v_{j+1}$. Let $\Delta S_i$ be a surface which intersects the vertex $v_{j+1}$ and is perpendicular to $l_i$, see figure 5. The size of $\Delta S_i$ corresponds to the initial lattice $\G_0$ in the sense that it spans an area corresponding to a side in a single cell.
The operator $L_{{\bf e}^a_i}$ should then, due to the Poisson bracket (\ref{Poisson}), be interpreted as a quantization of the flux variable $F^a_{\Delta S_i} $
\[
\mathrm{i}F^a_{\Delta S_i} \stackrel{\mbox{\tiny quantization}}{\longrightarrow} l_P^2 L_{{\bf e}^a_i}\;,
\]
where $l_P$ is the Planck length. It is important to realize that the inverse, densitised triad field involved in $F^a_{\Delta S_i} $ is located at the {\it endpoint} of $l_i$. Thus,  $F^a_{\Delta S_i} $ involves the quantity $\bar{E}_a^m(v_{j+1})$ through
\[
F^a_{\Delta S_i} =\int_{\Delta S_i} dx^2 \wedge dx^3 \bar{E}_a^1(v_{j+1})\;.
\]

\begin{figure}[t]
\begin{center}
\resizebox{!}{2.5cm}{
 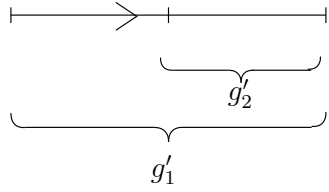}
\end{center}
\caption{An alternative partition of an edge into two.}
\end{figure}
Consider next the first subdivision of $l_i$ into two edges, which we denote $l'_i$ and $l'_{i+1}$. Thus, 
\[
\nabla(l_i)= \nabla(l'_i)\cdot \nabla(l'_{i+1})\;.
\] 
Also, denote the new vertex which subdivides $l_i$ by $v_{j+{1}/{2}}$. Now, the new copy of $G$ is associated to the first half of the line segment $l_i$, which means to $l_i'$. For notational simplicity, let us assume that this new copy of $G$  is the $(i+1)$'th copy of $G$ in $\ca_{\G_1}$ whereas the full line segment $l_i$ corresponds to the $i$'th copy of $G$. At first hand, it seems that the corresponding left-invariant vector fields $L_{{\bf e}^k_{i+1}}$ should be interpreted according to
\begin{equation}
\mathrm{i}F^a_{\Delta S_{i+1}} \stackrel{\mbox{\tiny quantization}}{\longrightarrow} l_P^2 L_{{\bf e}^a_{i+1}} \quad\quad\mbox{(first guess)}\;.
\label{first}
\end{equation}
However, this cannot be correct since $L_{{\bf e}^k_{i+1}}$ commutes with $\nabla(l_i)$ which belongs to the $i$'th copy of $G$. If equation (\ref{first}) should be correct then the commutator between $L_{{\bf e}^k_{i+1}}$ and $\nabla(l_i)$ should split up $\nabla(l_i)$ and insert a Lie-algebra generator at the new vertex $v_{j+1/2}$, since the edge $l_i$ intersects the surface $\Delta S_{i+1}$ at $v_{j+1/2}$. 
Instead, we find that relation (\ref{first}) obtains an additional term: 
\[
\mathrm{i}F^a_{\Delta S_{i+1}} \stackrel{\mbox{\tiny quantization}}{\longrightarrow}  l_P^2 L_{{\bf e}^k_{i+1}} +  l_P^2 R_{g_{i+1}{\bf e}^k_i g_{i+1}^{-1}}\;.
\]
Notice here that the triad field involved in $F^k_{\Delta S_{i+1}} $ is located at the new vertex $v_{j+1/2}$. If we had chosen a different partition of the line segment, see figure 6, then the left-invariant vector field corresponding to the new copy of $G$ would have an interpretation in terms of a flux variable and triad field located at $v_{j+1}$. Thus, the classical interpretation of $D$ distinguishes between the different modes of subdividing the line segment.

Notice also that the surfaces $\Delta S_i$ must shrink with each subdivision, in order to have one intersection point between the lattice and each surface. Thus, if we set the area of the initial surface equal to one, then the size of the surfaces decrease with subdivisions like
\begin{equation}
\vert \Delta S_i  \vert = 2^{-2n}\;.
\label{scaling}
\end{equation}

\begin{figure}[t]
\begin{center}
\resizebox{!}{4cm}{
 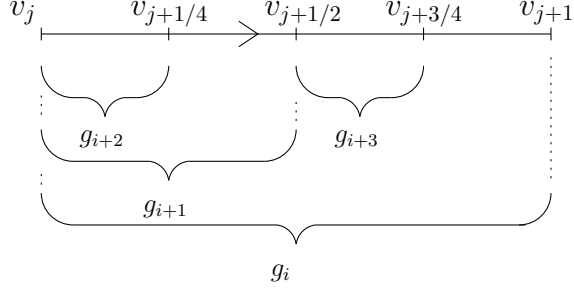}
\end{center}
\caption{Partition of an edge into four.}
\end{figure}
Consider the next subdivision of $l_i$ into four edges. The notation is as indicated in fig. 7. 
We find that the two new flux operators $F^a_{\Delta S_{i+2}}$ and $F^a_{\Delta S_{i+3}}$ have the following correspondences 
\[
\mathrm{i}F^a_{\Delta S_{i+2}} \stackrel{\mbox{\tiny quantization}}{\longrightarrow} l_P^2 L_{{\bf e}^a_{i+2}} +  l_P^2 R_{g_{i+2}{\bf e}^a_i g_{i+2}^{-1}}+ l_P^2  R_{g_{i+2}{\bf e}^a_{i+1} g_{i+2}^{-1}}\;,
\]
and
\[
\mathrm{i}F^a_{\Delta S_{i+3}} \stackrel{\mbox{\tiny quantization}}{\longrightarrow} l_P^2 L_{{\bf e}^a_{i+3}} + l_P^2  R_{g_{i+1}g_{i+3}{\bf e}^a_i g_{i+3}^{-1} g_{i+1}^{-1}}  \;.
\]
Once more, the particular subdivision of $l_i$ is singled out by this interpretation. If we had chosen the alternative subdivision of the edge into two, as pictured in figure 7, then this interpretation would not have been possible. 

There exist, however, at this level the possibility to choose the subdivision in figure 4.1. At this point of the analysis, there is no particular reason to chose between the two modes of subdivision pictured in figure 4, except perhaps that the subdivision in figure 4.2 is more symmetrical since new copies of $G$ are all assigned to edges of the same length. 

In general, at the $n$'th level of subdivision of $l_i$ we obtain the correspondence
\begin{equation}
F^k_{\Delta S_{i+s}} \stackrel{\mbox{\tiny quantization}}{\longrightarrow} l_P^2 L_{{\bf e}^k_{i+s}} +  l_P^2 \OO^k_{i+s-1} \;,
\label{xxx}
\end{equation}
where $\OO^k_{i+s-1}$ is a combination of twisted, right-invariant vector fields acting on the copies of $G$ assigned to edges which are situation "higher" in the inductive system of lattices. Put differently,  $\OO^k_{i+s-1}$ probes information which is more coarse grained relative to the line segment to which the $(i+s)$'th copy of $G$ is assigned.

In the following we shall ignore the correction terms $\OO^k_{i+s-1}$ when we apply relation (\ref{xxx}) to translate quantized quantities involving the Dirac type operator $D$ to their classical counterparts. The reason for this will become clear in the next section where we construct semi-classical states. These states have the property that any dependency on finite parts of the inductive system of lattices vanishes in the semi-classical limit.

In the limit of repeated subdivision of lattices we find that the semi-finite spectral triple $(\cb,\ch,D)$ encodes information tantamount to a representation of the Poisson bracket of general relativity. Thus, the triple carries information of the kinematical sector of quantum gravity. Clearly, the triple is based on a different set of variables than loop quantum gravity and hence the "representation" it encodes is different to the representation used there.

\section{Semiclassical analysis}

In this section we construct semi-classical states in $\ch$ and evaluate their expectation value of $D$.

\subsection{Coherent states on a Lie group}
We will first recall the results for coherent states on compact connected Lie groups that we are going to use. For simplicity we will only consider the case of most interest, namely $SU(2)$.
Let $\{e^a\}$ be a basis for $\mathfrak{su}(2)$. Given $g_0$ in $SU(2)$ and given three momenta (real numbers) $p^1,p^2,p^3$ there exist families $\phi_t\in L^2(SU(2))$ such that
$$ \lim_{t \to 0}\langle \phi^t, t L_{e^a}\phi^t \rangle=\mathrm{i}p^a\;,$$
and
$$\lim_{t \to 0}\langle \phi^t\otimes v, g\phi^t\otimes v \rangle=(v,g_0v)\;,$$
where $v \in M_2(\bbC)$, and $(,)$ denotes the inner product hereon.

Corresponding statements hold for operators of the type $$f(g)P(tL_{e^1},tL_{e^2},tL_{e^3}),$$  where $P$ is a polynomial in three variables, and $f$ is a smooth function on $SU(2)$, i.e.
$$ \lim_{t \to 0}\langle \phi^t,f(g)P(tL_{e^1},tL_{e^2},tL_{e^3}) \phi^t \rangle=f(g_0)P(\mathrm{i}p^1,\mathrm{i}p^2,\mathrm{i}p^3)\;.$$
This statement also carries over to symbols, i.e. functions on $T^*SU(2)$ with certain properties.

The construction of these states follows from work of Hall, see \cite{Hall1,Hall2}, and are more explicitely described in \cite{Thiemann:2000by} section 3.1. The states have further inportant physical properties, which we are however not going to use at the present stage of the analysis.

\subsection{Product states}
Let us consider the $n$'th level in a subdivision of lattices. We split the edges into $\{ l_i \}$, and $\{ l'_i\}$, where $\{ l_i\}$ denotes the edges appearing in the $n$'th subdivision but not in the $n-1$'th subdivision, and $\{ l'_i \}$ the rest. Define
$\phi^t_{l_i}$ to be the coherent state on $SU(2)$ such that
$$\lim_{t \to 0}\langle \phi^t_{l_i} \otimes v ,g \phi^t_{l_i}\otimes  v \rangle =  (v,h_{l_i}(A)v)\;,$$
and
$$\lim_{t \to 0}\langle \phi^t_{l_i} , tL_{e_i^a} ,\phi^t_{l_i}\rangle = 2^{-2n}\mathrm{i}E_a^m(v_{j+1})\;,$$
where $v \in M_2(\bbC)$; $v_{j+1}$ denotes the right endpoint of $l_i$, and the $m$ in the $E^m_a$ refers to the direction of $l_i$. The factor $2^{-2n}$ comes from the scaling (\ref{scaling}).
Furthermore define the states $\phi_{l'_i}$ by
$$\lim_{t \to 0}\langle \phi^t_{l'_i} \otimes v ,g \phi^t_{l'_i}\otimes  v \rangle =  (v,h_{l'_i}(A)v)\;,$$
and
$$\lim_{t \to 0}\langle \phi^t_{l'_i} , tL_{e_j^a} \phi^t_{l'_i}\rangle = 0\;.$$
Finally define $\phi^t_n$ to be the product of all these states as a state in $L^2(\ca_{\Gamma_n})$.

In the limit $n\rightarrow\infty$ these states produce the right expectation value on all loop operators in the infinite lattice.

\subsection{Semi-classical states: one copy of $G$}

\begin{figure}[t]
\begin{center}
\resizebox{!}{0.6cm}{
 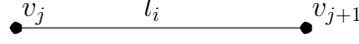}
\end{center}
\caption{A single edge.}
\end{figure}
We now proceed to construct semi-classical states in $\ch$. From here on we set $t=l_P^2$ and rescale the left-invariant vector fields in the Dirac type operator accordingly
$$
L_{{\bf e}_i^a}\rightarrow t L_{{\bf e}_i^a}\;.
 $$
The first step is to consider again a single edge.
Let $\psi(x)$ be a field on $\Sigma$. A priori, $\psi(x)$ can either be a two-spinor or a two-by-two matrix valued field. For reasons which shall become clear later, we choose the second option. 
Consider again an edge $l_i$ with endpoints $v_j$ and $v_{j+1}$, see figure 8. The states in $L^2(G,Cl(T^*G)\otimes M_2(\mathbb{C}))$ which we are interested in have the form\footnote{here we assume that $\psi(x)$ is matrix valued. If $\psi(x)$ was a two-spinor field then we  would instead consider the Hilbert space $L^2(G,Cl(T^*G)\otimes \mathbb{C}^2)$ and states therein.}
\begin{equation}
\Phi^t(l_i) = (g_i \psi(v_{j+1})+\mathrm{i} {\bf e}_i^a \s^a\psi(v_j) ) \phi^t_{l_i}\;,
\nonumber
\end{equation}
where the spinor field is evaluated at the endpoints of the edge $l_i$.
A straightforward computation gives the expectation value of $D$ on this state
\begin{eqnarray}
\lim_{t\rightarrow 0} \langle \bar{\Phi}^t \vert D \vert \Phi^t \rangle &=&
 2^{-2n}a_n  \big( -\bar{\psi}(v_j) \sigma^a E_a^m (\psi(v_{j+1})-\psi(v_j)) 
\nn\\&&
+     (\bar{\psi}(v_{j+1})-\bar{\psi}(v_j))  \sigma^a E_a^m \psi(v_j)  
\nonumber\\&& 
+ \bar{\psi}(v_j) \{\e A_m,\sigma^aE_a^m\}\psi(v_j)   \big)        \;,
\label{straightforward}
\end{eqnarray}
where we applied the expansion
\[
g= 1 + \e A_m + \co(\e^2)\;,
\]
with $\e=2^{-n}$. Also, the index $m$ denotes the direction of the edge $l_i$.

\subsection{Determining the sequence $\{a_n\}$}

Formula (\ref{straightforward}) indicates that the sequence $\{a_n\}$ of free parameters in $D$ plays a specific role in the semiclassical analysis. In particular, note the term
$(\psi(v_{j+1})-\psi(v_j)) $. If we consider the limit where the edge $l_i$ lies increasingly deep in the inductive system of graphs, then this term approaches
\begin{equation}
(\psi(v_{j+1})-\psi(v_j)) \rightarrow \pa_m \psi(v_j) dx^m\;,\quad (\mbox{no sum over $m$})\;,
\nonumber
\end{equation}
where $dx^m$ is the infinitesimal line segment, which goes as $2^{-n}$. 
Here $n$ denotes the level of subdivisions of graphs.
Thus, if we choose the sequence 
$$
a_n = 2^{3n}\;,
$$
then the expression (\ref{straightforward}) converges, when one considers edges of increasing depth in the inductive system of lattices, towards the quantity
\begin{eqnarray}
\lim_{n\rightarrow\infty} \lim_{t\rightarrow 0}   \langle \bar{\Phi}^t \vert D \vert \Phi^t \rangle &=& \bar{\psi}(v_j) \sigma^a E_a^m \nabla_m \psi(v_j)
- \nabla_m \bar{\psi}(v_{j})\sigma^a E_a^m \psi(v_j) \;,
\nn
\label{converge1}
\end{eqnarray}
(again, no sum over $m$) with $\nabla_m = \pa_m + A_m$.
This is the expectation value (in a point) of the self-adjoint operator
\begin{equation}
\sigma^a E_a^m \nabla_m + \nabla_m \sigma^a E_a^m\;.\quad (\mbox{no sum over $m$})\;
\nonumber
\end{equation}
Here, we applied what amounts to a partial integration (this will be justified shortly where an integral over $\Sigma$ emerges).

\subsection{Three copies of $G$}

\begin{figure}[t]
\begin{center}
\resizebox{!}{4cm}{
 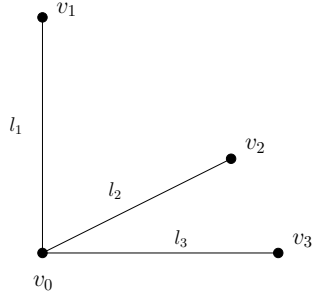}
\end{center}
\caption{Three edges, connected in one vertex.}
\end{figure}
Next, we consider instead three edges, denoted for simplicity by $l_1$, $l_{2}$, $l_{3}$, all leading out of the same vertex, with three copies of $G$ associated to them, correspondingly. First, consider the state
\begin{eqnarray}
\Phi^t(g_1,g_2,g_3) &=& \Big(
 {\bf e}_2^a {\bf e}_3^a g_1 \psi(v_1)
-
 {\bf e}_1^a  {\bf e}_3^a g_2\psi(v_2) 
 +
 {\bf e}_1^a {\bf e}_2^a g_3\psi(v_3) 
\nonumber\\&&
+
 \frac{\mathrm{i}}{5}  {\bf e}_1^a {\bf e}_2^b  {\bf e}_3^c  \big(  \d^{ab}\sigma^c +  \d^{ac}\sigma^b +  \d^{bc}\sigma^a    \big) \psi(v_0)\Big)
  \phi^t_{l_1}\phi^t_{l_2}\phi^t_{l_3}\;.
\label{ourstatesII}
\end{eqnarray}
where the enumeration of the vertices are show in figure 9. We find that the expectation value of $D$ on this state leads to the operator
\begin{equation}
\sigma^a E_a^m \nabla_m + \nabla_m \sigma^a E_a^m
\label{jj}
\end{equation}
in the limit where the edges $l_i$ lie increasingly deep in the inductive system of lattices. In equation (\ref{jj}) we now sum over $m$.

\subsection{Semiclassical states on $\overline{\ca} $}

To obtain semiclassical states on the full space $\overline{\ca} $ we need to prescribe a procedure to sum up the results for the individual copies of $G$, or rather, for vertices.

First, at the $n$'th level in the inductive system of lattices, where we have $n_{\G_n}$ copies of $G$, we write down the state
\begin{equation}
\Phi^t_n(\ca_{\G_n}) = \Delta_n\left(\sum_{v_j} \Psi_{v_j} \right)\phi_n^t \;,
\label{leveln}
\end{equation}
where $\Delta_n$ equals $2^{-3(n-1)/2}$. This will, in the limit taken below, converge to the Lebesque measure. Also, we define
\begin{eqnarray}
\Psi_{v_j} &=& 
 {\bf e}_{j_2}^a  {\bf e}_{j_3}^a g_{j_1}\psi(v_{j_1})
-
 {\bf e}_{j_1}^a  {\bf e}_{j_3}^a g_{j_2}\psi(v_{j_2}) 
+
 {\bf e}_{j_1}^a  {\bf e}_{j_2}^a g_{j_3} \psi(v_{j_3}) 
\nonumber\\&&
+
 \frac{\mathrm{i}}{10}  {\bf e}_{j_1}^a  {\bf e}_{j_2}^b  {\bf e}_{j_3}^c \big(  \d^{ab}\sigma^c +  \d^{ac}\sigma^b +  \d^{bc}\sigma^a    \big) \psi(v_j)
\;,
\label{ourstatesIII}
\end{eqnarray}
see figure 10. The sum in (\ref{leveln}) runs over a certain subclass of vertices in $\G_n$. At the $n$'th level, these vertices are the midpoints of the minimal cubes present at the $(n-1)$'th level. This discrimination between vertices admittedly appears to be somewhat arbitrary and it might be possible to take into account all edges. This, however, complicates matters. We shall return to this point in a later publication.
\begin{figure}[t]
\begin{center}
\resizebox{!}{4cm}{
 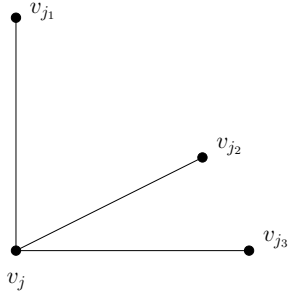}
\end{center}
\caption{Three more edges.}
\end{figure}

With (\ref{leveln}) we have a sequence $\{\Phi^t_n\}$ of states in $\ch$ and we can calculate the limit of the expectation value of $D$ on these states. We call this limit the continuum limit. We find
\begin{eqnarray}
\lim_{n\rightarrow\infty} \lim_{t\rightarrow 0}  \langle \bar{\Phi}^t_n \vert  D \vert \Phi^t_n  \rangle  
\nn\\
&&\hspace{-25mm}= \frac{1}{2}  \int_\Sigma d^3x \bar{ \psi}(x)(\sqrt{g}\sigma^a e_a^m\nabla_m + \nabla_m \sqrt{g}\sigma^a e^m_a  ) \psi(x)    \;.
\label{dex}
\end{eqnarray}
Thus, the sequence of states $\{\Phi^t_n\}$ defines a semi-classical limit where $D$, to lowest order, is a spatial Dirac operator on $\Sigma$. Notice that the integral in (\ref{dex}) is the invariant integral over $\Sigma$. The factor $\sqrt{g}$, where $g$ is the determinant of the spatial metric, comes from $\bar{E}_a^\m$. Here, however, it should be stressed that the emerging normalization of spinors $\psi(x)$ is {\it not} coordinate invariant. We shall comment on this below.

Note that the emergence of the integral in (\ref{dex}) crucially depends on the way the CAR algebra appear in the expression (\ref{ourstatesIII}).  Interestingly, the elements of the CAR algebra play the role of {\it localizers} in the construction.

\subsection{The Dirac Hamiltonian}

In equations (\ref{ourstatesII}) and (\ref{ourstatesIII}) we ignored certain degrees of freedom. To take these into account we modify the expression in equation (\ref{ourstatesIII}) to
\begin{eqnarray}
\tilde{\Psi}_{v_j} &=& 
 {\bf e}_{j_2}^a  {\bf e}_{j_3}^a g_{j_1}\psi(v_{j_1})
-
 {\bf e}_{j_1}^a  {\bf e}_{j_3}^a g_{j_2}\psi(v_{j_2}) 
+
 {\bf e}_{j_1}^a  {\bf e}_{j_2}^a g_{j_3} \psi(v_{j_3}) 
\nonumber\\&&
+
 \frac{\mathrm{i}}{20}  {\bf e}_{j_1}^a  {\bf e}_{j_2}^b  {\bf e}_{j_3}^c \big\{\big(  \d^{ab}\sigma^c +  \d^{ac}\sigma^b +  \d^{bc}\sigma^a    \big),M_{v_j}\big\} \psi(v_j)
\;,
\nonumber
\end{eqnarray}
where $M_{v_j}$ is an arbitrary self-adjoint two-by-two matrix. Write
$$
M_{v_j} = N(v_j) \mathds{1} +\mathrm{i} N^a(v_j) \sigma^a\;,
$$
where $N$ and $N^a$ are real fields on $\Sigma$, scalar and vectorial respectively. Finally, we let
$$
N^m = N^a e_a^m\;;
$$
we define
\begin{equation}
\tilde{\Phi}^t_n(\ca_{\G_n}) = \Delta_n\left(\sum_{v_j} \tilde{\Psi}_{v_j} \right)\phi_n^t\;,
\nonumber
\end{equation}
and
repeat the calculations leading to (\ref{dex}). We obtain
\begin{eqnarray}
\lim_{n\rightarrow\infty} \lim_{t\rightarrow 0}  \langle \bar{\tilde{\Phi}}^t_n \vert  D \vert \tilde{\Phi}^t_n  \rangle 
\nn\\
&&\hspace{-40mm}=  \int_\Sigma d^3x\bar{ \psi}(x) \left( \frac{1}{2}(\sqrt{g}N \sigma^a e_a^m\nabla_m + N\nabla_m \sqrt{g}\sigma^a e^m_a) +\mathrm{i}\sqrt{g} N^m\pa_m \right) \psi(x) 
\nonumber\\
&&\hspace{-40mm}+ 
\int_\Sigma d^3x\bar{ \psi}(x)\left(  \sqrt{g}N^m A_m + \frac{1}{2}( \pa_m  \sqrt{g}N^m) +  \frac{1}{2}(\pa_m N) \sqrt{g}\sigma^a e^m_a \right) \psi(x)
   \;.
\label{dex2}
\end{eqnarray}
Here, the first line is the principal part of the Dirac Hamiltonian in 3+1 dimensions. The second line contain additional zero-order terms. The fields $N$ and $N^m$ are seen to play the role of the lapse and shift fields respectively.

The additional zero-order terms appearing in (\ref{dex2}) are not identical to the zero-order terms in the Dirac Hamiltonian. This, however, is not to be expected since the Dirac Hamiltonian is not self-adjoint while the Dirac type operator is. For the reconstruction of the 4-metric only the principal part is used. We shall return to a discussion of the zeroth order part later. We believe that the correct treatment of the zeroth order terms can only be performed once the Wheeler-de-Witt constraint is formulated and implemented and thereby the freedom in choosing the foliation, i.e. the lapse and the shift fields, is eliminated.

This might also be a possible solution to another problem arising at this point. The norm of the semi-classical states $\tilde{\Phi}^t_n$ now depends on the lapse and shift fields,
$$
\lim_{n\rightarrow\infty} \lim_{t\rightarrow 0}  \langle \bar{\tilde{\Phi}}^t_n  \vert \tilde{\Phi}^t_n  \rangle 
=  \int_\Sigma d^3x\bar{ \psi}(x) \psi(x)\Omega(N,N^m)\;,
$$
where the function $\Omega(N,N^m)$ is readily computed.
This renders the interpretation of the semi-classical states as constituting the one-fermion states problematic as the induced scalar product is obviously not appropriate. Interestingly, however, the lapse and shift fields may be chosen such that $\Omega(N,N^m)=\sqrt{g}$. Thus, an appropriate choice of the time-coordinate restores the invariance of the norm under spatial diffeomorphisms. However, we are not aware of a compeling physical reason for such a choice of lapse and shift fields. Nevertheless, it might be conceivable that there is such a reason, as in quantum field theory, the one-particle space is not invariant under general coordinate changes. Thus, our restriction to one-particle states may well imply a restriction of the choice of coordinates.

The solution to the above problem might also lie in the construction of the states, i.e. it might be possible to modify the constrution of the semi-classical states such that the norm of the semi-classical spinors is automatically coordinate independent. We shall investigate this problem in future work.

Finally, disregarding lapse and shift fields, we should note that it would also be possible to remedy the deficiency of the missing $\sqrt{g}$ in the inner product by assigning the zero-order expectation value of the Halls coherent states to the non-densitised triad field and then adding, appropriately, the density in the semi-classical state. With this alteration the inner product of semi-classical states renders the correct inner product of spinors. However, this choice would spoil the interpretation of the left invariant vector fields as flux operators. 

Note that $\psi(x)$ takes values in $M_2(\bbC)$. In view of the action of the $\sigma^a$'s this can consistently be interpreted as a Dirac 4-spinor. The space spanned by these fields $\psi(x)$ can thus be interpreted as the space of solutions of the Dirac equation for the static 4-metric described by the 3-metric, the lapse and the shift fields (see \cite{Paschke}).

\section{ Discussion \& Outlook }

In this paper we have shown that to certain states for the previously constructed spectral triple over holonomy loops, one can associate gravitational and fermionic matter fields. This clearly indicates that one should interpret this model as describing quantized gravitational fields coupled to quantized matter fields.

To this concern we have constructed a small class of semi-classical states. 
Disregarding for the moment the open problem of identifying the correct scalar product, these semi-classical states can be interpreted as one-fermion states in a given foliation and given gravitational background field. We have identified the expectation value of the Dirac type operator of the spectral triple, in these semi-classical states, as the expectation value of the energy of the corresponding matter fields. This raises the question whether one can generally interpret the Dirac type operator as the energy operator for the matter fields present in the model. Thus, future work must clarify, first, whether there are many-particle fermionic states present in the model, and, of course, whether additional matter fields, for example photons, can be found. A consistent interpretation of the Dirac type operator then requires that it can also be interpreted as the energy of these states.

At the present state of the project the investigation of these issues is certainly within reach. 

A further strong indication that the model should be interpreted in terms of quantum gravity is the fact that it encodes information tantamount to a representation of the Poisson bracket of general relativity. This has been carefully analyzed for the first time in this paper and should therefore be seen as one of its central results.

All this being said, we should stress that our Hilbert space can only be viewed as the kinematical Hilbert space of quantum gravity. The Wheeler-de-Witt constraint has not been constructed nor implemented. In the construction above, this fact is nicely reflected by the appearence of the lapse and shift fields. Yet, as the Wheeler-de-Witt equation should in principle eliminate these unphysical degrees of freedom, the concreteness of their appearence raises the hope that our analysis may lead to a novel approach to the construction and implementation of the Hamiltonian constraint in quantum gravity.   \\

Apart from the physical interpretation of the model, the semi-classical analysis has also proven beneficial at a more technical level:
it turned out that the system of nested, cubic lattices, on which the semi-finite spectral triple is based, simply plays the role of a coordinate system. In particular, the lattices form the coordinate system already used to write down the Ashtekar variables and their Poisson bracket. This choice of background structure does, however, not imply lack of background invariance: there is no choice of background metric and the semi-clasical limit is coordinate independent. This shows that it is possible to recover the spatial symmetries with a countable system of lattices. Yet, it is an issue for future work to establish the full covariance of the model under change of the chosen coordinate system.

These observations are all based on the fact that any dependency on finite parts of the lattices vanishes in the limits (\ref{dex}) and (\ref{dex2}). That is, only the continuum limit contributes to the integrals in (\ref{dex}) and (\ref{dex2}). It is as if the lattices, which we have used to construct the spectral triple, disappear in this semi-classical limit.

Furthermore, the free parameters $\{a_n\}$, which appear in the Dirac type operator, play an important role in the semi-classical limit. A priori, this sequence is only required to diverge in order for the resolvent of the Dirac type operator to be compact. In the semi-classical limit, however, the sequence is identified as the inverse, infinitesimal volume element. This fixes the sequence.

We should stress that we only found states living on static 4-manifolds. This had to be expected since we interpret these states as one-particle states and it is well known in quantum field theory that such states would not exist on non-static space-times, e.g. in accelerating frames (which would be described by time-dependent lapse and shift fields). In the future it is certainly an interesting question whether one can find and describe semi-classical states which correspond to states of a quantized fermion field on a non-static space-time.

The application of the CAR algebra as a tool to form the local Riemann integral in equations (\ref{dex}) and (\ref{dex2}) is highly intriguing. It would certainly be very interesting and important to investigate the role played by the CAR algebra more thoroughly.

Moreover, the analysis in this paper is based on a real $SU(2)$ connection whereas the Ashtekar connection is complex. Geometrically, it is desirable to work with the original Ashtekar connection. One may speculate whether the complexity of the connection only appears in the semi-classical limit. If so, then one might exploit the techniques presented in this paper to obtain a complex connection via a doubling of the Hilbert space. \\

Immediate tasks to be addressed are: to compute quantum corrections for the semi-classical states in higher order of the Planck length; to investigate the operational interpretation of the loop algebra in the semi-classical states; to construct many particle states. Hopefully this will provide further evidence that the spectral triple over holonomy loops is a viable candidate for quantum gravity coupled to matter fields.    \\[1ex]

\noindent{\bf\large Acknowledgements}

\noindent 

J.A. and M.P. were supported by the SFB 478 grant  "Geometrische Strukturen in der Mathematik" of the Deutsche Forschungsgemeinschaft.

\end{document}

%% file: cube.pdf_t
\begin{picture}(0,0)%
\includegraphics{cube.pdf}%
\end{picture}%
\setlength{\unitlength}{4144sp}%
\begingroup\makeatletter\ifx\SetFigFont\undefined%
\gdef\SetFigFont#1#2#3#4#5{%
  \reset@font\fontsize{#1}{#2pt}%
  \fontfamily{#3}\fontseries{#4}\fontshape{#5}%
  \selectfont}%
\fi\endgroup%
\begin{picture}(1917,1651)(3316,-5750)
\put(3916,-5686){\makebox(0,0)[lb]{\smash{{\SetFigFont{12}{14.4}{\rmdefault}{\mddefault}{\updefault}$x^1$}}}}
\put(3736,-5236){\makebox(0,0)[lb]{\smash{{\SetFigFont{12}{14.4}{\rmdefault}{\mddefault}{\updefault}$x^2$}}}}
\put(3331,-5011){\makebox(0,0)[lb]{\smash{{\SetFigFont{12}{14.4}{\rmdefault}{\mddefault}{\updefault}$x^3$}}}}
\end{picture}%

%% file: AAAcubes.pdf_t
\begin{picture}(0,0)%
\includegraphics{AAAcubes.pdf}%
\end{picture}%
\setlength{\unitlength}{4144sp}%
\begingroup\makeatletter\ifx\SetFigFont\undefined%
\gdef\SetFigFont#1#2#3#4#5{%
  \reset@font\fontsize{#1}{#2pt}%
  \fontfamily{#3}\fontseries{#4}\fontshape{#5}%
  \selectfont}%
\fi\endgroup%
\begin{picture}(3894,1239)(3544,-5518)
\end{picture}%

%% file: AAPART.pdf_t
\begin{picture}(0,0)%
\includegraphics{AAPART.pdf}%
\end{picture}%
\setlength{\unitlength}{4144sp}%
\begingroup\makeatletter\ifx\SetFigFont\undefined%
\gdef\SetFigFont#1#2#3#4#5{%
  \reset@font\fontsize{#1}{#2pt}%
  \fontfamily{#3}\fontseries{#4}\fontshape{#5}%
  \selectfont}%
\fi\endgroup%
\begin{picture}(7229,1857)(661,-2593)
\put(1846,-916){\makebox(0,0)[lb]{\smash{{\SetFigFont{12}{14.4}{\rmdefault}{\mddefault}{\updefault}{$G$}%
}}}}
\put(2701,-916){\makebox(0,0)[lb]{\smash{{\SetFigFont{12}{14.4}{\rmdefault}{\mddefault}{\updefault}{$G$}%
}}}}
\put(2386,-2356){\makebox(0,0)[lb]{\smash{{\SetFigFont{12}{14.4}{\rmdefault}{\mddefault}{\updefault}{$G$}%
}}}}
\put(2611,-1681){\makebox(0,0)[lb]{\smash{{\SetFigFont{12}{14.4}{\rmdefault}{\mddefault}{\updefault}{$P$}%
}}}}
\put(6301,-1591){\makebox(0,0)[lb]{\smash{{\SetFigFont{12}{14.4}{\rmdefault}{\mddefault}{\updefault}$g'_2$}}}}
\put(6751,-2086){\makebox(0,0)[lb]{\smash{{\SetFigFont{12}{14.4}{\rmdefault}{\mddefault}{\updefault}$g'_1$}}}}
\end{picture}%

%% file: AATWO.pdf_t
\begin{picture}(0,0)%
\includegraphics{AATWO.pdf}%
\end{picture}%
\setlength{\unitlength}{4144sp}%
\begingroup\makeatletter\ifx\SetFigFont\undefined%
\gdef\SetFigFont#1#2#3#4#5{%
  \reset@font\fontsize{#1}{#2pt}%
  \fontfamily{#3}\fontseries{#4}\fontshape{#5}%
  \selectfont}%
\fi\endgroup%
\begin{picture}(9072,2383)(-509,-1394)
\put(271, 29){\makebox(0,0)[lb]{\smash{{\SetFigFont{12}{14.4}{\rmdefault}{\mddefault}{\updefault}$g_3$}}}}
\put(676,-376){\makebox(0,0)[lb]{\smash{{\SetFigFont{12}{14.4}{\rmdefault}{\mddefault}{\updefault}$g_2$}}}}
\put(1621,-1276){\makebox(0,0)[lb]{\smash{{\SetFigFont{12}{14.4}{\rmdefault}{\mddefault}{\updefault}$g_1$}}}}
\put(1126,-826){\makebox(0,0)[lb]{\smash{{\SetFigFont{12}{14.4}{\rmdefault}{\mddefault}{\updefault}$g_4$}}}}
\put(-494,839){\makebox(0,0)[lb]{\smash{{\SetFigFont{12}{14.4}{\rmdefault}{\mddefault}{\updefault}1.}}}}
\put(4501,839){\makebox(0,0)[lb]{\smash{{\SetFigFont{12}{14.4}{\rmdefault}{\mddefault}{\updefault}2.}}}}
\put(5671,-421){\makebox(0,0)[lb]{\smash{{\SetFigFont{12}{14.4}{\rmdefault}{\mddefault}{\updefault}$g_2$}}}}
\put(6571,-1321){\makebox(0,0)[lb]{\smash{{\SetFigFont{12}{14.4}{\rmdefault}{\mddefault}{\updefault}$g_1$}}}}
\put(7021,-781){\makebox(0,0)[lb]{\smash{{\SetFigFont{12}{14.4}{\rmdefault}{\mddefault}{\updefault}$g_4$}}}}
\put(5221, 74){\makebox(0,0)[lb]{\smash{{\SetFigFont{12}{14.4}{\rmdefault}{\mddefault}{\updefault}$g_3$}}}}
\end{picture}%

%% file: AADS.pdf_t
\begin{picture}(0,0)%
\includegraphics{AADS.pdf}%
\end{picture}%
\setlength{\unitlength}{4144sp}%
\begingroup\makeatletter\ifx\SetFigFont\undefined%
\gdef\SetFigFont#1#2#3#4#5{%
  \reset@font\fontsize{#1}{#2pt}%
  \fontfamily{#3}\fontseries{#4}\fontshape{#5}%
  \selectfont}%
\fi\endgroup%
\begin{picture}(4886,2004)(2731,-6463)
\put(6616,-5371){\makebox(0,0)[lb]{\smash{{\SetFigFont{12}{14.4}{\rmdefault}{\mddefault}{\updefault}{$l_{i+1}$}%
}}}}
\put(3376,-5371){\makebox(0,0)[lb]{\smash{{\SetFigFont{12}{14.4}{\rmdefault}{\mddefault}{\updefault}{$l_i$}%
}}}}
\put(2746,-5326){\makebox(0,0)[lb]{\smash{{\SetFigFont{12}{14.4}{\rmdefault}{\mddefault}{\updefault}{$v_{j}$}%
}}}}
\put(5221,-5281){\makebox(0,0)[lb]{\smash{{\SetFigFont{12}{14.4}{\rmdefault}{\mddefault}{\updefault}{$v_{j+1}$}%
}}}}
\put(4816,-4831){\makebox(0,0)[lb]{\smash{{\SetFigFont{12}{14.4}{\rmdefault}{\mddefault}{\updefault}{$\Delta S_i$}%
}}}}
\end{picture}%

%% file: PART2.pdf_t
\begin{picture}(0,0)%
\includegraphics{PART2.pdf}%
\end{picture}%
\setlength{\unitlength}{4144sp}%
\begingroup\makeatletter\ifx\SetFigFont\undefined%
\gdef\SetFigFont#1#2#3#4#5{%
  \reset@font\fontsize{#1}{#2pt}%
  \fontfamily{#3}\fontseries{#4}\fontshape{#5}%
  \selectfont}%
\fi\endgroup%
\begin{picture}(2051,1210)(6019,-1934)
\put(7426,-1366){\makebox(0,0)[lb]{\smash{{\SetFigFont{12}{14.4}{\rmdefault}{\mddefault}{\updefault}$g'_2$}}}}
\put(6931,-1861){\makebox(0,0)[lb]{\smash{{\SetFigFont{12}{14.4}{\rmdefault}{\mddefault}{\updefault}$g'_1$}}}}
\end{picture}%

%% file: AAAA.pdf_t
\begin{picture}(0,0)%
\includegraphics{AAAA.pdf}%
\end{picture}%
\setlength{\unitlength}{4144sp}%
\begingroup\makeatletter\ifx\SetFigFont\undefined%
\gdef\SetFigFont#1#2#3#4#5{%
  \reset@font\fontsize{#1}{#2pt}%
  \fontfamily{#3}\fontseries{#4}\fontshape{#5}%
  \selectfont}%
\fi\endgroup%
\begin{picture}(3852,2125)(4711,-944)
\put(6571,-871){\makebox(0,0)[lb]{\smash{{\SetFigFont{12}{14.4}{\rmdefault}{\mddefault}{\updefault}$g_i$}}}}
\put(5671,-421){\makebox(0,0)[lb]{\smash{{\SetFigFont{12}{14.4}{\rmdefault}{\mddefault}{\updefault}$g_{i+1}$}}}}
\put(5221, 74){\makebox(0,0)[lb]{\smash{{\SetFigFont{12}{14.4}{\rmdefault}{\mddefault}{\updefault}$g_{i+2}$}}}}
\put(5581,974){\makebox(0,0)[lb]{\smash{{\SetFigFont{14}{16.8}{\rmdefault}{\mddefault}{\updefault}$v_{j+1/4}$}}}}
\put(6526,974){\makebox(0,0)[lb]{\smash{{\SetFigFont{14}{16.8}{\rmdefault}{\mddefault}{\updefault}$v_{j+1/2}$}}}}
\put(7381,974){\makebox(0,0)[lb]{\smash{{\SetFigFont{14}{16.8}{\rmdefault}{\mddefault}{\updefault}$v_{j+3/4}$}}}}
\put(8326,974){\makebox(0,0)[lb]{\smash{{\SetFigFont{14}{16.8}{\rmdefault}{\mddefault}{\updefault}$v_{j+1}$}}}}
\put(4726,974){\makebox(0,0)[lb]{\smash{{\SetFigFont{14}{16.8}{\rmdefault}{\mddefault}{\updefault}$v_j$}}}}
\put(7021, 74){\makebox(0,0)[lb]{\smash{{\SetFigFont{12}{14.4}{\rmdefault}{\mddefault}{\updefault}$g_{i+3}$}}}}
\end{picture}%

%% file: one22.pdf_t
\begin{picture}(0,0)%
\includegraphics{one22.pdf}%
\end{picture}%
\setlength{\unitlength}{4144sp}%
\begingroup\makeatletter\ifx\SetFigFont\undefined%
\gdef\SetFigFont#1#2#3#4#5{%
  \reset@font\fontsize{#1}{#2pt}%
  \fontfamily{#3}\fontseries{#4}\fontshape{#5}%
  \selectfont}%
\fi\endgroup%
\begin{picture}(2363,350)(-52,786)
\put( 46,929){\makebox(0,0)[lb]{\smash{{\SetFigFont{14}{16.8}{\rmdefault}{\mddefault}{\updefault}$v_j$}}}}
\put(2296,929){\makebox(0,0)[lb]{\smash{{\SetFigFont{14}{16.8}{\rmdefault}{\mddefault}{\updefault}$v_{j+1}$}}}}
\put(991,929){\makebox(0,0)[lb]{\smash{{\SetFigFont{14}{16.8}{\rmdefault}{\mddefault}{\updefault}$l_i$}}}}
\end{picture}%

%% file: three.pdf_t
\begin{picture}(0,0)%
\includegraphics{three.pdf}%
\end{picture}%
\setlength{\unitlength}{4144sp}%
\begingroup\makeatletter\ifx\SetFigFont\undefined%
\gdef\SetFigFont#1#2#3#4#5{%
  \reset@font\fontsize{#1}{#2pt}%
  \fontfamily{#3}\fontseries{#4}\fontshape{#5}%
  \selectfont}%
\fi\endgroup%
\begin{picture}(2730,2871)(-329,446)
\put(136,3134){\makebox(0,0)[lb]{\smash{{\SetFigFont{14}{16.8}{\rmdefault}{\mddefault}{\updefault}$v_1$}}}}
\put(1936,1829){\makebox(0,0)[lb]{\smash{{\SetFigFont{14}{16.8}{\rmdefault}{\mddefault}{\updefault}$v_2$}}}}
\put(2386,929){\makebox(0,0)[lb]{\smash{{\SetFigFont{14}{16.8}{\rmdefault}{\mddefault}{\updefault}$v_3$}}}}
\put(-89,524){\makebox(0,0)[lb]{\smash{{\SetFigFont{14}{16.8}{\rmdefault}{\mddefault}{\updefault}$v_0$}}}}
\put(-314,2009){\makebox(0,0)[lb]{\smash{{\SetFigFont{12}{14.4}{\rmdefault}{\mddefault}{\updefault}$l_1$}}}}
\put(631,1379){\makebox(0,0)[lb]{\smash{{\SetFigFont{12}{14.4}{\rmdefault}{\mddefault}{\updefault}$l_2$}}}}
\put(1261,929){\makebox(0,0)[lb]{\smash{{\SetFigFont{12}{14.4}{\rmdefault}{\mddefault}{\updefault}$l_3$}}}}
\end{picture}%

%% file: curt.pdf_t
\begin{picture}(0,0)%
\includegraphics{curt.pdf}%
\end{picture}%
\setlength{\unitlength}{4144sp}%
\begingroup\makeatletter\ifx\SetFigFont\undefined%
\gdef\SetFigFont#1#2#3#4#5{%
  \reset@font\fontsize{#1}{#2pt}%
  \fontfamily{#3}\fontseries{#4}\fontshape{#5}%
  \selectfont}%
\fi\endgroup%
\begin{picture}(2505,2904)(-104,437)
\put(136,3134){\makebox(0,0)[lb]{\smash{{\SetFigFont{14}{16.8}{\rmdefault}{\mddefault}{\updefault}$v_{j_1}$}}}}
\put(-89,524){\makebox(0,0)[lb]{\smash{{\SetFigFont{14}{16.8}{\rmdefault}{\mddefault}{\updefault}$v_j$}}}}
\put(1936,1829){\makebox(0,0)[lb]{\smash{{\SetFigFont{14}{16.8}{\rmdefault}{\mddefault}{\updefault}$v_{j_2}$}}}}
\put(2386,929){\makebox(0,0)[lb]{\smash{{\SetFigFont{14}{16.8}{\rmdefault}{\mddefault}{\updefault}$v_{j_3}$}}}}
\end{picture}%

%% file: NewPaper_IV.bbl
\begin{thebibliography}{99}





\bibitem{Aastrup:2005yk}
  J.~Aastrup and J.~M.~Grimstrup,
  ``Spectral triples of holonomy loops,''
  Commun.\ Math.\ Phys.\  {\bf 264} (2006) 657
  [arXiv:hep-th/0503246].


\bibitem{Aastrup:2006ib}
  J.~Aastrup and J.~M.~Grimstrup,
  ``Intersecting Connes noncommutative geometry with quantum gravity,''
  Int.\ J.\ Mod.\ Phys.\  A {\bf 22} (2007) 1589
  [arXiv:hep-th/0601127].




\bibitem{Aastrup:2008wa}
  J.~Aastrup, J.~M.~Grimstrup and R.~Nest,
  ``On Spectral Triples in Quantum Gravity I,''
  Class.\ Quant.\ Grav.\  {\bf 26} (2009) 065011
  [arXiv:0802.1783 [hep-th]].



\bibitem{Aastrup:2008wb}
  J.~Aastrup, J.~M.~Grimstrup and R.~Nest,
  ``On Spectral Triples in Quantum Gravity II,''
  J.\ Noncommut.\ Geom.\  {\bf 3} (2009) 47
  [arXiv:0802.1784 [hep-th]].


\bibitem{Aastrup:2008zk}
  J.~Aastrup, J.~M.~Grimstrup and R.~Nest,
  ``A new spectral triple over a space of connections,''
  Commun.\ Math.\ Phys.\  {\bf 290} (2009) 389
  [arXiv:0807.3664 [hep-th]].



\bibitem{Aastrup:2009ux}
  J.~Aastrup, J.~M.~Grimstrup and R.~Nest,
  ``Holonomy Loops, Spectral Triples $\&$ Quantum Gravity,'' to appear in Class. Quant. Grav.,
  [arXiv:0902.4191 [hep-th]].



\bibitem{ConnesBook}
A.~Connes,
``Noncommutative Geometry,'' Academic Press, 1994.



\bibitem{Connes:1996gi}
A.~Connes,
``Gravity coupled with matter and the foundation of non-commutative
geometry,''
Commun.\ Math.\ Phys.\  {\bf 182} (1996) 155
[arXiv:hep-th/9603053].






\bibitem{Thiemann:2001yy}
  T.~Thiemann,
  ``Introduction to modern canonical quantum general relativity,''
  [arXiv:gr-qc/0110034].




\bibitem{Rovelli:2004tv}
  C.~Rovelli,
  ``Quantum gravity,''
Cambridge, UK: Univ. Pr. (2004) 455 p.



\bibitem{Ashtekar:2004eh}
  A.~Ashtekar and J.~Lewandowski,
  ``Background independent Quantum Gravity: A status report,''
  Class.\ Quant.\ Grav.\  {\bf 21} (2004) R53
  [arXiv:gr-qc/0404018].


\bibitem{Hall1}
Brian C. Hall. "The Segal-Bargmann "coherent state" transform for compact Lie groups". J. Funct. Anal., 122(1):103-151, 1994.

\bibitem{Hall2}
Brian C. Hall. "Phase space bounds for quantum mechanics on a compact Lie group". Comm. Math. Phys., 184(1):233-250, 1997.




\bibitem{Ashtekar:1993wf}
A.~Ashtekar and J.~Lewandowski,
``Representation theory of analytic holonomy C* algebras,''
[arXiv:gr-qc/9311010].




\bibitem{Ashtekar:1994wa}
  A.~Ashtekar and J.~Lewandowski,
 ``Differential geometry on the space of connections via graphs and projective
  limits,''
  J.\ Geom.\ Phys.\  {\bf 17} (1995) 191
  [arXiv:hep-th/9412073].



\bibitem{Ashtekar:1996eg}
  A.~Ashtekar and J.~Lewandowski,
  ``Quantum theory of geometry. I: Area operators,''
  Class.\ Quant.\ Grav.\  {\bf 14} (1997) A55
  [arXiv:gr-qc/9602046].





\bibitem{Flori:2008nw}
  C.~Flori and T.~Thiemann,
  ``Semiclassical analysis of the Loop Quantum Gravity volume operator: I. Flux
  Coherent States,''
  [arXiv:0812.1537 [gr-qc]].






\bibitem{Chamseddine:1991qh}
A.~H.~Chamseddine and A.~Connes,
``Universal formula for noncommutative geometry actions: Unification of
gravity and the standard model,''
Phys.\ Rev.\ Lett.\  {\bf 77} (1996) 4868.
 
\bibitem{Chamseddine:1996rw}
A.~H.~Chamseddine and A.~Connes,
``A universal action formula,''
[arXiv:hep-th/9606056].

\bibitem{Chamseddine:1996zu}
A.~H.~Chamseddine and A.~Connes,
``The spectral action principle,''
Commun.\ Math.\ Phys.\ {\bf 186} (1997) 731
[arXiv:hep-th/9606001].




\bibitem{Chamseddine:2006ep}
  A.~H.~Chamseddine, A.~Connes and M.~Marcolli,
  ``Gravity and the standard model with neutrino mixing,''
  [arXiv:hep-th/0610241].

\bibitem{Chamseddine:2007hz}
  A.~H.~Chamseddine and A.~Connes,
  ``Why the Standard Model,''
  [arXiv:0706.3688 [hep-th]].

\bibitem{Chamseddine:2007ia}
  A.~H.~Chamseddine and A.~Connes,
  ``A Dress for SM the Beggar,''
  [arXiv:0706.3690 [hep-th]].







\bibitem{Ashtekar:1986yd}
  A.~Ashtekar,
  ``New Variables for Classical and Quantum Gravity,''
  Phys.\ Rev.\ Lett.\  {\bf 57} (1986) 2244.

\bibitem{Ashtekar:1987gu}
  A.~Ashtekar,
  ``New Hamiltonian Formulation of general relativity,''
  Phys.\ Rev.\  D {\bf 36} (1987) 1587.






\bibitem{Thiemann:2000by}
  T.~Thiemann and O.~Winkler,
  ``Gauge field theory coherent states (GCS). IV: Infinite tensor product  and
  thermodynamical limit,''
  Class.\ Quant.\ Grav.\  {\bf 18} (2001) 4997
  [arXiv:hep-th/0005235].




\bibitem{Paschke}
M. Paschke and T. Kopf.
"A spectral quadruple for de Sitter space",
J. Math. Phys. {\bf 43}, 818 (2002)
[arXiv:math-ph/0012012].











\end{thebibliography}
